\documentclass{article}

\usepackage[preprint]{neurips_2025}

\usepackage[utf8]{inputenc}
\usepackage[T1]{fontenc}
\usepackage{hyperref}
\usepackage{url}
\usepackage{booktabs}
\usepackage{amsfonts}
\usepackage{nicefrac}
\usepackage{microtype}
\usepackage{xcolor}

\usepackage{amsmath}
\usepackage{amssymb}
\usepackage{graphicx}
\graphicspath{ {./figures/} }
\usepackage{comment}
\usepackage{subcaption}

\title{When Independent Gaussian Models Break Down: Characterizing Regime-Dependent Modeling Failures in $\phi^4$ Theory}

\author{%
  Anish Bhat\thanks{Equal contribution.} \\
  University of California, San Diego \\
  \url{a2bhat@ucsd.edu}
  \And
  Ryo Ide\footnotemark[1] \\
  University of California, San Diego \\
  \url{ride@ucsd.edu}
  \And
  Zihan Zhao \\
  University of California, San Diego \\
  \url{ziz078@ucsd.edu}}

\begin{document}

\maketitle


\begin{abstract}
In practical physical systems, modeling assumptions of Gaussianity and basis independence break down due to self-interactions. We study a specific instance of one-dimensional $\phi^4$ theory on a lattice, analyzing how the interaction strength and system size jointly affect the marginal and joint distributions of frequency-based representation of the field (i.e., Fourier modes). We find that models relying on Gaussian and independent Fourier modes fail primarily due to structured dependencies rather than marginal non-Gaussianity, since individual modes become approximately Gaussian despite mode coupling growing with size. Based on this, we identify three distinct regimes that delineate where traditional methods remain effective and where more expressive models are needed. Our results provide a computationally simple diagnostic to establish when Gaussian models are insufficient, and derive structural requirements that motivate future nonlinear model designs.
\end{abstract}


\section{Introduction}
Methods that assume an independent basis for modeling physical systems decompose complex systems into simple components, making them a crucial tool for machine learning modeling of physical phenomena. However, this relies on the assumption that the physical system can be decomposed into modes that act independently, which is often violated in real systems. 

One such system is the one-dimensional quartic interaction (1D $\phi^4$ Theory), a type of interaction in quantum field theory. 
Self-interactions are deliberately introduced in the modeling action principle by adding $\lambda\phi^4$ where $\phi$ is the field and $\lambda$ is the coupling constant that determines interaction strength. 

While certain bases, such as a Fourier basis, diagonalize non-interacting field theories and imply statistical independence between modes, interactions in $\phi^{4}$ theory can break this assumption through mode coupling. Currently, it is unknown to what extent Fourier-based methods can still accurately describe $\phi^{4}$, at what scales, and when independence-based Gaussian representations become insufficient. 
 
Our main contributions are: (1) testing the limits of current methods based on independent representations; (2) testing independence-agnostic data-driven methods; (3) identifying the coupling regimes where methods fail; (4) identifying properties that future nonlinear models may need to express.


\section{Problem setup}
To address these questions, we compare Fourier-based Gaussian models with data-driven baselines using samples from a Markov Chain-Monte Carlo (MCMC) simulation of $\phi^{4}$ theory, which is well established in this subject \cite{PhysRevLett, Hasenbusch_1999, Albergo2019}. Our full experimental setup is available at \url{https://anonymous.4open.science/r/phi4-PAI-2026-Stanford-F12B/README.md}.

\paragraph{Field Distribution.}
We consider a 1D lattice scalar field $\phi \in \mathbb{R}^N$ with periodic boundary conditions. Our MCMC implementation targets the following Boltzmann distribution $p(\phi) \propto e^{-S[\phi]}$, where the action is $S[\phi] = \sum_{i=1}^{N} \left[
\frac{1}{2}(\phi_{i+1} - \phi_i)^2
+ \frac{1}{2} m^2 \phi_i^2
+ \lambda \phi_i^4
\right]$ \cite{jansen2011phi4}. $S[\phi]$ encodes the energy cost of a field configuration; the Boltzmann distribution weights configurations by how energetically favorable they are, with $\lambda$ controlling the strength of self-interaction.

Note that when $\lambda=0$, the distribution $p(\phi)$ can be fully factorized into a pure Gaussian. In general, increasing $\lambda$ increases interaction strength.

\paragraph{Sampling.}
Samples are generated using the Metropolis-Hastings MCMC algorithm on a field with coupling constants  $\lambda \in \{0.1, 1.0, 5.0, 10.0, 20.0, $ $50.0, 100.0\}$. 



\paragraph{Observables.}
The power spectrum is computed via the discrete Fourier transform $\tilde{\phi}_k = \mathcal{F}[\phi]_k$, where $P(k) = \mathbb{E}[|\tilde{\phi}_k|^2]$~\cite{oppenheim2010dtsp}.

\section{Models and baselines}

\paragraph{Fourier Model.}
As a baseline, we use a Fourier-based method traditionally used to model free theories without self-interactions, which we apply here to $\phi^4$ Theory. We assume that Fourier modes are independent Gaussian variables $\tilde{\phi}_k \sim \mathcal{N}(0, \sigma^2_k),$ with $\sigma_k^2$ estimated from data. Samples are generated by independently sampling each mode and applying the inverse Fourier transform. 

\paragraph{Principal Component Analysis Model.}
As a data-driven baseline, we use PCA to find $z = U^\top (\phi - \bar{\phi})$, where $U$ are the eigenvectors of $\Sigma = U \Lambda U^T$~\cite{abdi2010pca}. 

\paragraph{Full-Gaussian Model.}
As an independence-agnostic baseline, we fit $\tilde{\phi}\sim N(0,\Sigma)$ directly to data, testing how far second-order Gaussian behavior persists as interaction increases.

\paragraph{Masked Autoregressive Flow. }
For a nonlinear baseline, we train a masked autoregressive flow (MAF) on Fourier-space samples \cite{papamakarios2017maf}, learning a nonlinear transformation from the base Gaussian in order to represent arbitrary joint structure including 4th-order inter-mode dependencies.

\section{Experimental design}

\paragraph{Normalized Coupling Metric}
First, we define spectral energy $E_k = |\tilde{\phi}_k|^2$. $E_k$ is a random variable over the sample ensemble, so $\mathrm{Cov}(E_k, E_{k'})$ is estimated across samples. Second, computing covariance $\Sigma^{(E)} = \mathrm{Cov}(E_k, E_{k'})$, we define the normalized coupling metric as $C = \frac{\|\Sigma^{(E)} - \mathrm{diag}(\Sigma^{(E)})\|_F}{
\|\Sigma^{(E)}\|_F}.$ For jointly Gaussian variables, statistical independence implies zero pairwise covariance (i.e., $C = 0$). $C$ therefore measures how much the spectral energy variables violate independence, with $C=1$ indicating that off-diagonal structure dominates. Since $E_k = |\tilde{\phi}_k|^2$ is quadratic in $\tilde{\phi}_k$, the covariance $\mathrm{Cov}(E_k, E_{k'})$ involves fourth-order moments of the field, making $C$ a fourth-order statistic.

\paragraph{Spectral Error Metric}

To measure model performance, we define normalized spectral error $\epsilon = \frac{\sum_k |\hat{P}(k) - P(k)|}{\sum_k P(k)}$, where $P(k)$ is the empirical power spectrum and $\hat{P}(k)$ is the predicted power spectrum from generated samples.


\paragraph{Excess Kurtosis.}
To quantify deviations from Gaussianity in marginal distributions, we compute the excess kurtosis $K(x)$ for position-space variables $\phi_i$ and Fourier modes $\tilde{\phi}_k$ \cite{nist}. 
A negative (positive) $K$ indicates lighter (heavier) tails, providing a simple diagnostic of marginal non-Gaussianity. 

\paragraph{Kullback-Leibler Divergence.}
To measure deviations from Gaussian structure, we compute the Kullback–Leibler (KL) Divergence $D_{\mathrm{KL}}(p\|q)$ between the empirical distribution $p(x)$ and a Gaussian reference $q(x)$ with matched mean and covariance \cite{Shalizi}. We estimate this quantity on 1D marginals (per-site and per-mode), quantifying marginal non-Gaussianity more sensitively than kurtosis. Unlike joint metrics such as the coupling metric $C$, $D_{\mathrm{KL}}(p\|q)$ does not capture mode coupling, allowing us to explicitly separate marginal effects from inter-mode dependencies.


\section{Results}
\begin{figure}[hbt]
    \centering
    \begin{subfigure}[b]{0.32\linewidth}
        \centering
        \includegraphics[width=\linewidth]{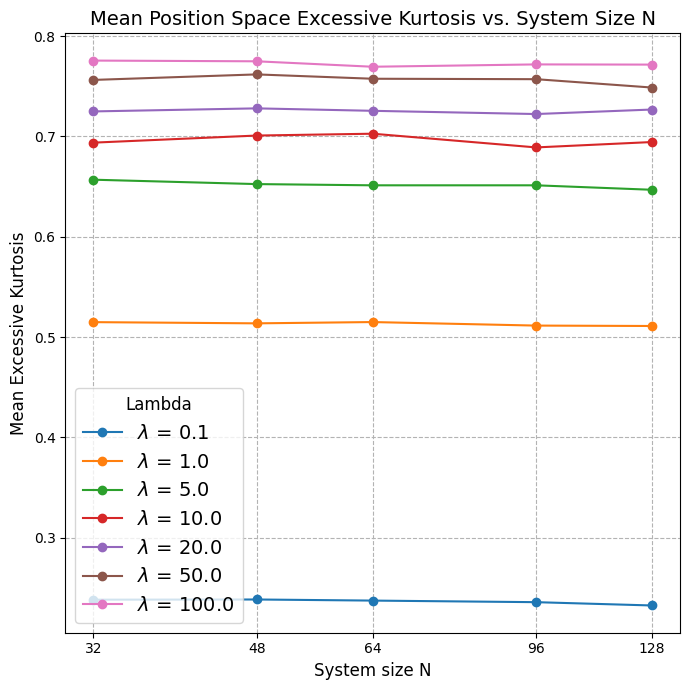}
        \caption{Mean $K_\text{position}$ Magnitude vs. $N$}
        \label{fig:1a}
    \end{subfigure}
    \begin{subfigure}[b]{0.32\linewidth}
        \centering
        \includegraphics[width=\linewidth]{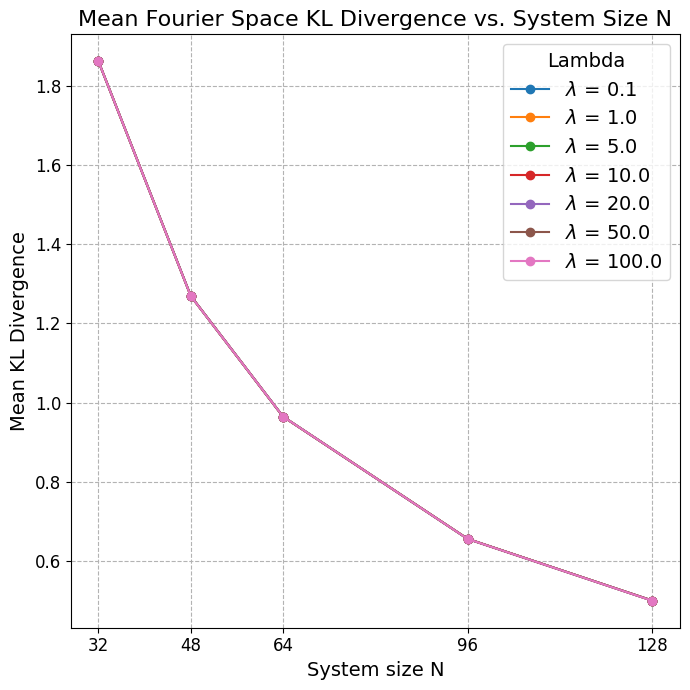}
        \caption{Mean $D_{\text{KL, Fourier}}$ vs. $N$}
        \label{fig:1b}
    \end{subfigure}
    \caption{
    ~\ref{fig:1a}: Empirically, for $\lambda=0.1$, $K_{\text{position}}\approx -0.23$, indicating near-Gaussian 
    behavior. Largely independent of $N$, it saturates at $-0.77$, near the theoretical limit of $-0.81$ given by standard $\phi^4$ theory.
    ~\ref{fig:1b}: Empirically, $D_{\text{KL, Fourier}}$ gradually decreases to $D_{\text{KL, Fourier}}\approx 0.4$. Similarly, $K_{\text{Fourier}}$ approaches $\approx0.14$ with $N$, likewise indicating near-Gaussian marginals. Note that curves overlap.
    }
    \label{fig:gaussianity}
\end{figure}

\paragraph{Gaussianity Analysis.}
First, we examine marginal Gaussianity in position space via excess kurtosis $K_{\text{position}}$. At weak coupling, near-Gaussian behavior is observed (see Figure \ref{fig:1a}). 
$K_{\text{position}}$ is largely invariant with system size $N$, indicating that local statistics are determined primarily by interaction strength. 
While local marginals become strongly non-Gaussian at large $\lambda$, they are effectively fixed once the system is sufficiently large. This suggests that Gaussianity-agnostic density models (e.g., Variational Autoencoders (VAEs) or Normalizing Flows) may better model local field distributions by capturing this non-Gaussian structure \cite{DBLP:journals/corr/abs-1906-02691, Kobyzev_2021, Caselle2022esc}.

\begin{figure}[hbt]
    \centering
    \begin{subfigure}[b]{0.32\linewidth}
        \centering
        \includegraphics[width=\linewidth]{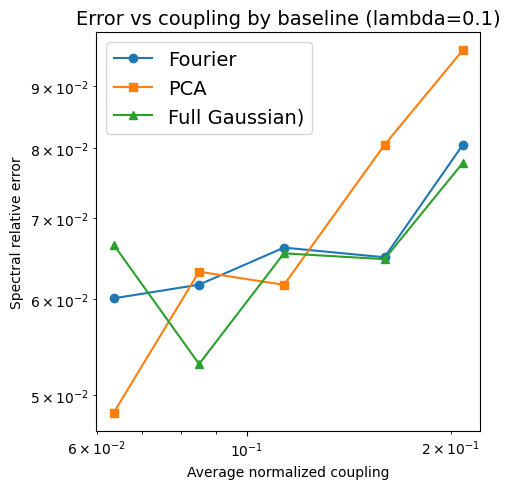}
        \caption{$\lambda=0.1$}
        \label{fig:2a}
    \end{subfigure}
    \begin{subfigure}[b]{0.32\linewidth}
        \centering
        \includegraphics[width=\linewidth]{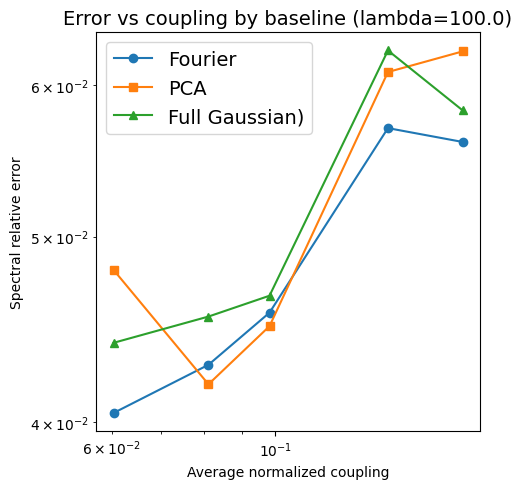}
        \caption{$\lambda=100.0$}
        \label{fig:2b}
    \end{subfigure}
        \begin{subfigure}[b]{0.34\linewidth}
        \centering
        \includegraphics[width=\linewidth]{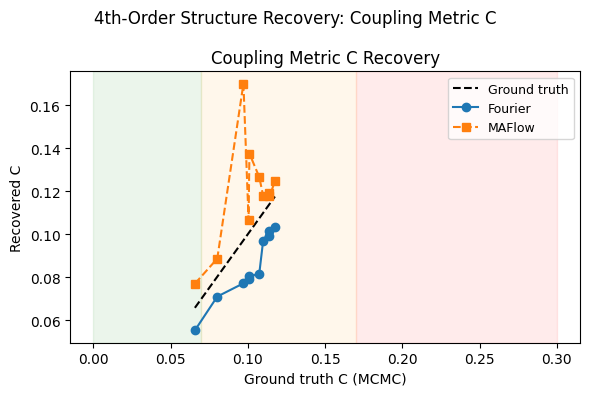}
        \caption{MAF vs. Fourier}
        \label{fig:2c}
    \end{subfigure}
    
    \caption{
    \ref{fig:2a}-\ref{fig:2b} Spectral relative error vs. normalized coupling across all baselines. Similar trends are observed for spectral error with respect to system size $N$. This is consistent with coupling increasing with $N$: across all $\lambda$, $C$ ranges from $C(N=32)\approx0.06$ to $C(N=128)\approx0.2$. \ref{fig:2c} Recovered coupling C vs. ground truth C across all baselines. The Fourier baseline systematically underestimates C by construction; the MAF recovers higher coupling values but overshoots, outperforming Fourier in half of intermediate-regime settings.
    }
    \label{fig:baselines}
\end{figure}

In Fourier space, both excess kurtosis $K_{\text{Fourier}}$ and KL divergence $D_{\text{KL, Fourier}}$ decrease with $N$ and approach stable values with nearly identical behavior across all $\lambda$ (see Figure \ref{fig:1b}). This indicates that individual Fourier modes become approximately Gaussian in distribution. While this trend is consistent with Central-Limit-Theorem-type behavior arising from averaging over many degrees of freedom, the convergence occurs even as interactions remain present. Thus, marginal Gaussianity alone does not imply statistical independence between modes.
Despite the increasingly Gaussian Fourier marginals, we observe that the spectral error $\epsilon$ grows monotonically with coupling $C$ (see Figure \ref{fig:2a}--\ref{fig:2b}). Since performance is evaluated across Gaussian-family models here, this result suggests that accurate marginal statistics are insufficient to guarantee accurate joint modeling. Instead, performance appears increasingly limited by the presence of structured inter-mode dependencies as coupling grows. This motivates future investigation of models representing higher-order joint structure, such as VAEs and Normalizing Flows \cite{Albergo2021intro, Kanwar2020}.

\paragraph{Regime-Dependent Behavior.}
We identify three empirical regimes based on the measured coupling value $C$ (see Figure \ref{fig:2a}--\ref{fig:2b}): (i) a weak-coupling regime ($C \lesssim 0.07$) where spectral error remains small and independent representations perform well; (ii) an intermediate-coupling regime ($0.07 \lesssim C \lesssim 0.17$) where increasing coupling is accompanied by a corresponding increase in spectral error; and (iii) a strong-coupling regime ($C \gtrsim 0.17$) where both coupling and spectral error saturate. In this regime, $C \in (0.17,0.18)$ and $\epsilon \in (0.05,0.06)$ for most tested settings. Although spectral error is a second-order observable, its monotonic relationship with the fourth-order coupling metric $C$ suggests that it remains sensitive to the same interaction-driven structure captured by $C$. These observations suggest that $C$ separates regimes more directly than $\lambda$ or $N$ alone.

To examine whether recovering the joint 4th-order structure is possible, we evaluate the MAF on $C$ (see Figure \ref{fig:2c}). While the Fourier baseline systematically underestimates $C$ (0.088 vs. ground truth 0.105 in the intermediate regime), the MAF consistently overestimates $C$ (0.128), outperforming Fourier in only half of intermediate-regime settings.

\section{Limitations}

\paragraph{Restricted Physical Setup.}
We study a 1D lattice with periodic boundary conditions and a fixed bare mass $m^2$. Higher dimensions, alternative boundary conditions, and the symmetry-breaking phase transition of $\phi^4$ as a function of $m^2$ may alter the regime structure identified here. A joint sweep over $(m^2, \lambda)$ is left for future work.

\paragraph{Diagnostic and Model Scope.}
The metric $C$ is a 4th-order statistic and does not constrain higher cumulants ($\geq 6$) that may matter at strong coupling. Our MAF baseline probes 4th-order structure but exhibits training instability, so we cannot distinguish whether residual strong-coupling error is fundamental or capacity-limited. We do not estimate the MCMC sampling noise floor at fixed ($\lambda$, N), so we cannot rule out that the observed saturation of $\epsilon$ is partially a noise artifact.


\section{Conclusion}

We identify three coupling regimes in 1D $\phi^4$ theory and find a separation between marginal simplicity and joint complexity: Fourier modes become roughly Gaussian marginally as $N$ grows, while joint coupling $C$ grows by a factor of ${\sim}3$. Gaussian baselines fit marginals well by construction, suggesting that residual error increasingly reflects unexpressed joint higher-order structure.

Our MAF results show that nonlinear models can capture inter-mode coupling that Gaussian baselines miss by design, though instability in training limits consistent recovery. Our results suggest that successful models require that they (i) preserve approximate marginal Gaussianity of Fourier modes, and (ii) capture 4th-order joint cumulants between modes. These requirements explain why fully factorized Gaussian families, full-covariance Gaussian models, and linear bases leave residual error in this regime, which further motivates coupling-layer normalizing flows and nonlinear latent-variable models. While demonstrated in 1D $\phi^4$ theory, the coupling diagnostic $C$ is expected to arise in any system where tunable self-interactions compete with a natural independent-mode decomposition.

\section{Acknowledgement}
This work was performed using the National Research Platform Nautilus HyperCluster. We would like to thank the researchers at the Duarte Lab at UC San Diego for their mentorship and support. Additionally, we would like to thank the organizers of PAI 2026 and Stanford University for giving us the opportunity to present our research.

\newpage
\bibliographystyle{plainnat}
\bibliography{bibliography}

\end{document}